# An efficient way to model complex magnetite:

# assessment of SCC-DFTB against DFT.

Hongsheng Liu[1], Gotthard Seifert[2], Cristiana Di Valentin[1]*

[1] Dipartimento di Scienza dei Materiali, Università di Milano-Bicocca, via R. Cozzi 55, I-20125 Milano, Italy

[2] Theoretische Chemie, Technische Universität Dresden, 01062, Dresden, Germany

**Abstract**

Magnetite has attracted increasing attention in recent years due to its promising and diverse applications in biomedicine. Theoretical modelling can play an important role in understanding magnetite-based nanomaterials at the atomic scale for a deeper insight into the experimental observations. However, calculations based on density functional theory (DFT) are too costly for realistically large models of magnetite nanoparticles. Classical force field methods are very fast but lack of precision and of the description of electronic effects. Therefore, a cheap and efficient quantum mechanical simulation method with comparable accuracy than DFT is highly desired. Here, a less computational demanding DFT-based method, i.e. self-consistent charge density functional tight-binding (SCC-DFTB), is adopted to investigate magnetite bulk and low-index (001) surface with newly proposed parameters for Fe-O interactions. We report that SCC-DFTB with on-site Coulomb correction provides results in quantitatively comparable agreement with those obtained by DFT+U and hybrid functional methods. Therefore, SCC-DFTB is valued as an efficient and reliable method for the description magnetite. This assessment will promote SCC-DFTB computational studies on magnetite-based nanostructures that attract increasing attention for medical applications.

# I. INTRODUCTION

Magnetite ($Fe_3O_4$), a ferrimagnet with a high Curie temperature of about 850 K, has attracted increasing attention in recent years due to its promising applications in biomedicine, including drug delivery, magnetic resonance imaging (MRI), magnetic hyperthermia, bioseparation and biosensor.[1,2,3,4,5,6,7] At room temperature, magnetite crystallizes in an inverse spinel structure with oxygen anions arranged in a slightly distorted face centered cubic lattice and iron atoms occupying tetrahedral and octahedral interstitial sites. Both $Fe^{2+}$ and $Fe^{3+}$ exist in magnetite in a ratio of 1 : 2, with tetrahedral sites occupied by $Fe^{3+}$ and octahedral sites occupied by an equal number of $Fe^{2+}$ and $Fe^{3+}$.[8]

In general, the applications of magnetite in biomedicine are based on the attachment of ligands to $Fe_3O_4$ nanoparticles, which is clearly a surface issue. Therefore, a lot of effort has been devoted to the investigation of magnetite surfaces, as summarized in a recent review.[9] For the $Fe_3O_4(001)$ surface, considering the $(\sqrt{2} \times \sqrt{2})R45°$ reconstruction that appears in experiments[10,11,12,13,14,15], R. Pentcheva and co-workers, proposed a distorted bulk truncation (DBT) model based on DFT calculations,[16] which is thermodynamically very stable.[16,17] However, the Pendry reliability factor ($R_P$) for the low-energy electron diffraction (LEED) investigation was somewhat poor ($R_P = 0.34$)[18] and the DBT model could not explain the site preference of Au adatoms deposited on the $Fe_3O_4(001)$ surface.[19] In 2014 Bliem et al. proposed a reconstructed surface model (SCV), that is a B layer terminated $Fe_3O_4(001)$ surface, with an extra interstitial $Fe_{Tet}$ atom in the second layer replacing two $Fe_{Oct}$ atoms that are removed from the third layer per $(\sqrt{2} \times \sqrt{2})R45°$ unitcell.[20] The SCV model agrees well with the surface X-ray diffraction[21] and shows a much better agreement with experimental LEED *IV* ($R_P = 0.125$)[20] compared with the DBT model ($R_P = 0.34$)[18]. In addition, the

SCV model shows, at the DFT+U level of theory, a higher thermodynamic stability than DBT over the entire range of oxygen chemical potentials accessible in experimentally conditions and can well explain the site preference of Au adatoms when deposited on $Fe_3O_4(001)$ surface.[20] In contrast, the structure of $Fe_3O_4(111)$ surface is not yet clear because there are two competitive terminations that are close in energy at oxygen poor conditions and are difficult to distinguish by STM.[9] The $Fe_3O_4(110)$ surface is also not well understood up to now, though several experimental[22,23,24] and computational[25,26,27] works have been performed. Besides the termination of surfaces, molecular adsorption, such as that of water,[28,29,30,31,32,33,34,35,36] carbon monoxide,[37] formic acid,[38] methanol,[39] polyvinyl alcohol,[36] polyethylene glycol[36] and so on, has also been widely explored.

However, in contrast with the large amount of fundamental works on magnetite bulk and surfaces, magnetite nanoparticles, which are the mostly interesting structures in nanomedicine, have only been studied experimentally and in reference to their synthesis and medical applications.[1, 2, 3, 4, 5, 6, 7] Only few computational studies exist on the organic molecules adsorption on magnetite nanoparticles, based on force-field methods,[40,41,42,43] which intrinsically lack the description of electronic and magnetic effects and of the chemical processes at the interface. First-principles calculations may play an important role in the achievement of an atomic level understanding of magnetite nanoparticles structure and properties. However, standard DFT methods fail in the detailed description of electronic and magnetic properties of magnetite.[44] More sophisticated approaches are required to catch proper effects, such as DFT+U or hybrid functionals, which, however, are very expensive and time consuming. Therefore, the scientific community is in urgent need for a cheaper method but still sufficiently accurate for such a complex material as magnetite.

The self-consistent charge density functional tight binding (SCC-DFTB)[45] is an approximate quantum chemical method derived from DFT based on a second-order expansion of the DFT total energy expression. SCC-DFTB, which is a fast and efficient quantum mechanical simulation method, provides information about electronic structure properties, which is not available from classical force fields. In the past few years, SCC-DFTB has been successfully exploited to investigate various types of systems, including semiconductor (Si/SiO$_2$ interfaces [46]), hybrid inorganic-organic systems (gold-thiolate compounds,[47] organic molecules on GaAs(l00) surface[48]), low dimensional materials (MoS$_2$ nanotubes,[49] chrysotile nanotubes[50] and MoS$_2$ fullerenes[51]), transition metal and relative metalorganic complexes [52] and even transition metal oxides (TiO$_2$ bulk, surfaces and nanoparticles).[53,54,55,56,57,58,59,60,61]

In this study, we extend the available DFTB parametrizations for the description of magnetite, proposing a parametrization of the Fe–O interactions. The performance of these new parameters is assessed through the comparison of SCC-DFTB data for several structural, electronic and magnetic properties of magnetite bulk and (001) surface with those obtained by DFT+U and hybrid functional calculations in previous works.[16,20,34,44] Our study shows that results by SCC-DFTB with the on-site Coulomb correction[62] (DFTB+U) are in quantitatively agreement with those by DFT+U and hybrid functional methods. The excellent performance of DFTB on magnetite provides an alternative and cheap approach to the modelling of realistically large magnetite nanoparticles, which are the crucial systems in nanomedicine.

## II. METHODS

The SCC-DFTB method is an approximated DFT-based method that derives from the second-order expansion of the Kohn-Sham total energy in DFT with respect to the electron density fluctuations. The SCC-DFTB total energy can be defined as:

$$E_{tot} = \sum_i^{occ} \varepsilon_i + \frac{1}{2} \sum_{\alpha,\beta}^N \gamma_{\alpha\beta} \Delta q_\alpha \Delta q_\beta + E_{rep} \qquad (1),$$

where, the first term is the sum of the one-electron energies $\varepsilon_i$ coming from the diagonalization of an approximated Hamiltonian matrix. $\Delta q_\alpha$ and $\Delta q_\beta$ are the induced charges on the atoms α and β, respectively, and $\gamma_{\alpha\beta}$ is a Coulombic-like interaction potential. $E_{rep}$ is a short-range pairwise repulsive potential. More details about the SCC-DFTB method can be found in Refs. 45, 63, and 64. From now on, DFTB will be used as shorthand for SCC-DFTB.

All the DFTB calculations were performed by the DFTB+ package.[65] For the Fe-Fe and O-O interactions we used the "trans3d-0-1" and "mio-1-1" set of parameters as reported previously.[45,66] For the Fe-O interactions, we first tried the Slater-Koster files from "trans3d-0-1". Then we refitted the repulsive part to improve its performance. To properly deal with the strong correlation effects among Fe 3d electrons, DFTB+U[62] with an effective U-J value of 3.5 eV was adopted according to our previous work on magnetite bulk and (001) surface based on DFT.[34,44] The convergence criterion of $10^{-4}$ a.u. for force was used during geometry optimization and the convergence threshold on the self-consistent charge (SCC) procedure was set to be $10^{-5}$ a.u.

For bulk magnetite, the conventional cell containing 32 oxygen atoms and 24 iron atoms was adopted and the k points generated by the Monkhorst−Pack scheme were chosen to be 6 ×6 ×6. For the Fe₃O₄(001) surface, two structural models were considered, DBT and SCV, as we used in our previous paper[34]. An inversion symmetric slab with 17 atomic layers was adopted for both DBT and SCV structures (see Figure 2). A ($\sqrt{2} \times \sqrt{2}$)R45° supercell for the (001) surface was used for the DBT and SCV models containing 124 ($Fe_{52}O_{72}$) and 122 ($Fe_{50}O_{72}$) atoms, respectively. In the z direction a vacuum of more than 12 Å was introduced to avoid the spurious interaction between periodic images.

Five atomic layers in the middle are kept fixed to the bulk position and the other layers are fully relaxed during the geometry optimization. For the surfaces, the K points mesh was $6 \times 6 \times 1$.

For all the details on the DFT+U and HSE calculations on magnetite bulk and surface systems, the reader is referred to our previous studies where they are reported.[34,44]

## III. RESULTS AND DISCUSSION

The geometry of bulk magnetite was first fully relaxed with both DFTB and DFTB+U methods with the Slater-Koster files from "trans3d-0-1" and "mio-1-1" libraries, labeled as DFTB-lib and DFTB+U-lib, respectively. The optimized lattice parameters by different methods together with the experimental value are listed in Table 1. The lattice parameters given by DFTB-lib and DFTB+U-lib are much larger than the experimental value[67] and values from PBE, PBE+U, HSE, reported previously.[44] The bad agreement is not surprising because the Fe-O Slater-Koster files were fitted for bioinorganic systems.[66]

**Table 1.** Lattice parameter (a) for the conventional cell of bulk magnetite obtained by different methods.

| Methods | DFTB-lib | DFTB+U-lib | DFTB-refitted | DFTB+U-refitted | PBE[44] | PBE+U[44] | HSE[44] | EXP.[67] |
|---------|----------|------------|---------------|-----------------|---------|-----------|---------|----------|
| a (Å) | 8.763 | 8.898 | 8.314 | 8.488 | 8.380 | 8.491 | 8.389 | 8.394 |

To improve the performance of DFTB on magnetite, we refitted the repulsive part of the Fe-O Slater-Koster file (the $E_{rep}$ term in formula (1)). To do this, DFT calculations with Perdew Burke Ernzerhof (PBE) functional were performed for the conventional cell of magnetite using the plane-wave-based Quantum ESPRESSO package[68]. The detailed setup can be found in our previous paper for bulk magnetite.[44] A series of calculations for bulk magnetite with different lattice parameters were performed with both DFTB-lib and PBE to obtain the energy dependence on the Fe-O distance. For each lattice parameter, the coordinates for all the atoms in the conventional cell were fully relaxed with PBE. The energy difference between electronic energy from DFTB-lib and binding energy from PBE can be viewed as the repulsive energy in formula (1). The $E_{rep}$ is a pairwise repulsive potential. However, there are two kinds of Fe-O bonds in bulk magnetite, $Fe_{Oct}$-O bond and $Fe_{Tet}$-O bond. Therefore, to represent the repulsive energy as a function of Fe-O distance, a special average (R) of $Fe_{Tet}$-O and $Fe_{Oct}$-O distances ($R_{Tet}$ and $R_{Oct}$) was computed as:

$$R = \sqrt{\frac{4R_{Tet}^2 R_{Oct}^2}{R_{Tet}^2 + 3R_{Oct}^2}} \qquad (2).$$

The calculated $E_{rep}$ per Fe-O pair as a function of the average Fe-O distance is shown in Figure 1 (black points). This was fitted by a polynomial:[63]

$$E_{rep}(R) = \begin{cases} \sum_{n=2} d_n (R_c - R)^n, & for\ R < R_c, \\ 0, & otherwise. \end{cases} \qquad (3),$$

where, $R_c$ is the cutoff radius and $d_n$ are the coefficients. The best fitting is given by $R_c = 4.3$ Bohr. The fitted $E_{rep}$ as a function of the average Fe-O distance are plotted in Figure 1 (red line).

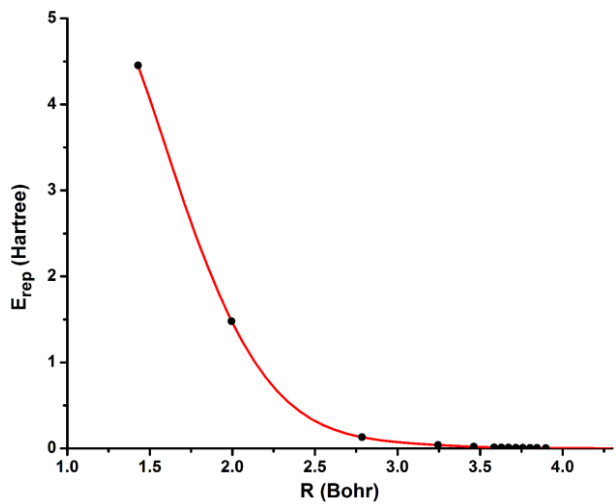

**Figure 1.** The repulsive energy as a function of the average Fe-O distance in bulk magnetite. The black points are obtained by calculations and the red line is the plotting of the fitted $E_{rep}$.

We used the refitted Fe-O Slater-Koster file to fully relax the geometry of bulk magnetite again with both DFTB and DFTB+U methods, labeled as DFTB-refitted and DFTB+U-refitted, respectively. DFTB+U-refitted results in a lattice parameter similar to those from PBE+U and HSE, only about 1.1% larger than the experimental value.[67] In contrast, DFTB-refitted underestimates the lattice parameter with respect to the experimental value, similarly to the PBE method. Therefore, DFTB+U with the refitted Fe-O Slater-Koster file can well describe the structural properties of bulk magnetite and will be adopted hereafter. From now on, we will refer to DFTB+U-refitted as DFTB+U.

We now discuss how the DFTB+U method describes the electronic properties. The distribution of $Fe^{2+}$ and $Fe^{3+}$ at different lattice sites is shown in Table 2. All the Fe ions at tetrahedral sites are 3+, in accordance with PBE+U and HSE.[44] However, DFTB+U cannot reproduce the charge disproportionation for the bulk magnetite, i.e. all the Fe ions at octahedral sites have the same valence states, which can be viewed as $Fe^{2.5+}$. This is because the high structural symmetry of bulk magnetite prevents an approximated DFT-based method to correctly describe the symmetry breaking of the wavefunction that is required to distinguish $Fe^{3+}$ and $Fe^{2+}$ at geometrically equivalent octahedral sites. Such delicate effect can be observed at the DFT+U and hybrid functionals level of theory.[44] All the Fe ions at tetrahedral sites couple antiferromagnetically with Fe ions at octahedral sites (see Table 2). DFTB+U slightly overestimates the atomic magnetic moment compared with PBE+U and HSE. Note that the total magnetic moment for the conventional cell is fixed to be 32 $\mu_B$ for DFTB+U calculations according to the value given by PBE+U.

**Table 2.** Number of Fe ions at tetrahedral and octahedral sites with different valence ($N_{Tet}^{3+}$, $N_{Oct}^{3+}$, $N_{Oct}^{2+}$), charge difference between $Fe_{Oct}^{3+}$ and $Fe_{Oct}^{2+}$ ($\Delta Q$), magnetic moment of Fe ions at tetrahedral and octahedral sites ($m_{Tet}$, $m_{Oct}^{3+}$, $m_{Oct}^{2+}$) in bulk magnetite derived from different methods: DFTB+U, PBE+U and HSE.

| | DFTB+U | PBE+U[44] | HSE[44] |
|---|---|---|---|
| $N_{Tet}^{3+}$ | 8 | 8 | 8 |
| $N_{Oct}^{3+}$ | 16 | 8 | 8 |
| $N_{Oct}^{2+}$ | | 8 | 8 |
| $\Delta Q$ (e) | 0.0 | 0.27 | 0.32 |
| $m_{Tet}$ ($\mu_B$) | -4.53 | -3.96 | -4.21 |
| $m_{Oct}^{3+}$ ($\mu_B$) | 4.11 | 4.06 | 4.27 |
| $m_{Oct}^{2+}$ ($\mu_B$) | | 3.62 | 3.79 |

The projected density of states (PDOS) on the *d* states of different Fe ions is calculated by DFTB+U (Figure 2a). The PDOS shows a half-metal character with some spin down states from $Fe_{Oct}$ at the Fermi level. This agrees well with those obtained by PBE+U (Figure 2b) and HSE (Figure 2c) with symmetry constrained wavefunction.[44] The lack of band gap for the spin down channel is because of the uniform valence state of Fe ions ($Fe^{2.5+}$) at octahedral sites, as discussed previously.[44] A band gap between $d$ states of Fe at octahedral and tetrahedral sites exists for the spin up channel, which is similar to the results from PBE+U and HSE.

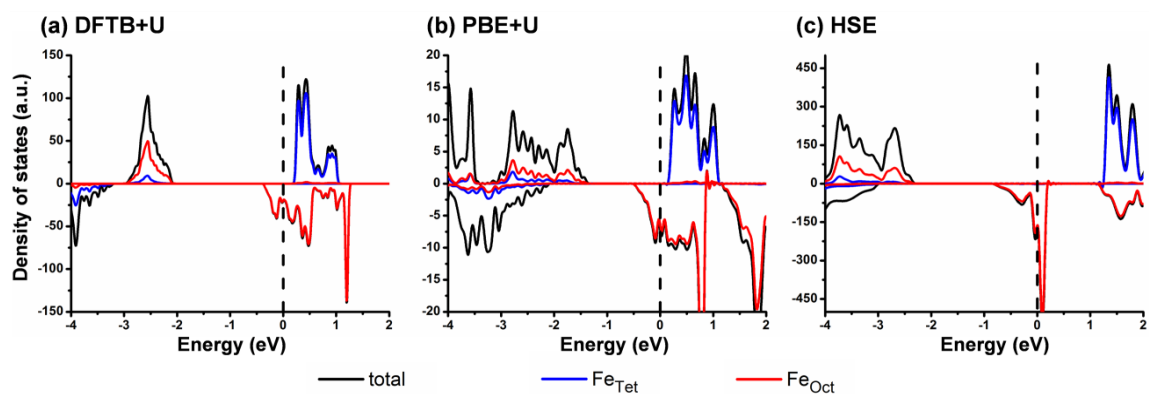

**Figure 2.** Projected density of states on the d states of different Fe ions in bulk magnetite calculated with (a) DFTB+U, (b) PBE+U and (c) HSE. Legend of colors is on the bottom. Data for (b) and (c) is taken from our previous paper.[44]

To check the performance of DFTB+U on $Fe_3O_4$ surfaces, the two proposed models (DBT and SCV, as discussed in the introduction) of the well-studied (001) surface were fully relaxed with DFTB+U. For the DBT surface, no undulation along the surface $Fe_{Oct}$ rows was observed (Figure 3a). In contrast, the undulations in the SCV surface model are very pronounced (Figure 3b). These observations agree well with previous results based on DFT+U and HSE calculations.[20,34] For both DBT and SCV surfaces, the distance between the first two layers is largely compressed with respect to the bulk value (0.828 and 0.812 Å vs. 1.061 Å, respectively), while the distance between the second and the third layers is only slightly changed. The relaxation in the surface layers is also in accordance with that obtained by DFT+U and HSE methods.[34] Therefore, DFTB+U can well describe the structural details of magnetite surfaces.

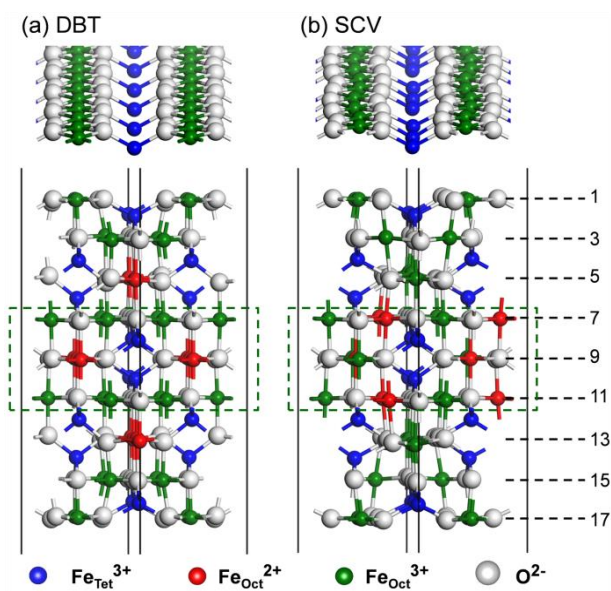

**Figure 3.** Top and side views of optimized atomic structures of the (a) DBT and (b) SCV $Fe_3O_4(001)$ surface models by DFTB+U. Layers are numbered on the right. Atomic layers in the dashed rectangles are kept fixed in the bulk positions during atomic relaxation.

**Table 3.** Numbers of Fe ions with different valence at different sites ($N_{Tet}^{3+}$, $N_{Oct}^{3+}$, $N_{Oct}^{2+}$), charge difference between $Fe_{Oct}^{3+}$ and $Fe_{Oct}^{2+}$ ($\Delta Q$), magnetic moment of Fe ions at tetrahedral and octahedral sites ($m_{Tet}$, $m_{Oct}^{3+}$, $m_{Oct}^{2+}$) in $Fe_3O_4(001)$ surfaces derived from different methods: DFTB+U, PBE+U and HSE.

| | DBT | | | SCV | | |
|---|---|---|---|---|---|---|
| | DFTB+U | PBE+U[34] | HSE[34] | DFTB+U | PBE+U[34] | HSE[34] |
| $N_{Tet}^{3+}$ | 16 | 16 | 16 | 18 | 18 | 18 |
| $N_{Oct}^{3+}$ | 24 | 24 | 24 | 26 | 26 | 26 |
| $N_{Oct}^{2+}$ | 12 | 12 | 12 | 6 | 6 | 6 |
| $\Delta Q$ (e) | 0.63 | 0.35 | 0.37 | 0.65 | 0.34 | 0.38 |
| $m_{Tet}$ ($\mu_B$) | -4.51~-4.57 | -3.96~-3.98 | -4.20~-4.24 | -4.53~-4.58 | -3.92~-3.99 | -4.18~-4.24 |
| $m_{Oct}^{3+}$ ($\mu_B$) | 4.50~4.61 | 4.05~4.09 | 4.29~4.30 | 4.49~4.59 | 4.04~4.09 | 4.26~4.33 |
| $m_{Oct}^{2+}$ ($\mu_B$) | 3.71~3.76 | 3.62~3.67 | 3.75~3.84 | 3.70~3.74 | 3.60~3.66 | 3.76~3.83 |

Fe ions with different valence at tetrahedral and octahedral sites are marked with different colors and shown in Figure 3. DFTB+U method is capable of reproducing the charge disproportionation of Fe ions at the octahedral sites. The surface cut to build the slab models reduces the symmetry with respect to the bulk and makes it easier for the DFTB+U method to break the wavefunction symmetry and thus distinguish $Fe^{2+}$ and $Fe^{3+}$ at different sites. In the outmost two layers of the DBT model all $Fe_{Oct}$ are $Fe^{3+}$ due to the low Fe:O stoichiometry in surface layers. On the contrary, all $Fe_{Oct}$ in layers 5 and 13 (hereafter called as "deep surface layers") are $Fe^{2+}$. In contrast, all $Fe_{Oct}$ in the fully relaxed layers are $Fe^{3+}$ for SCV model (Figure 3b) as a consequence of even lower Fe:O ratio in the surface layers due to the Fe vacancies present in the third layer. The high $Fe^{3+}$ density at surface layers is consistent with XPS measurements.[20] For both DBT and SCV surfaces, in those layers that are kept fixed during atomic relaxation (from $7^{th}$ layer to $11^{th}$ layer), $Fe^{2+}$ and $Fe^{3+}$ ions alternate, as observed in the bulk phase.[44] Iron cations at tetrahedral sites are always $Fe^{3+}$, whether at the surface or in the pseudo-bulk layers. The distribution of $Fe^{2+}$ and $Fe^{3+}$ in the surface slabs just described for DFTB+U calculations agrees well with that obtained by PBE+U and HSE.[34] In addition, the numbers of Fe ions with different valence at different sites are exactly the same as those obtained by PBE+U and HSE (listed in Table 3). Therefore, DFTB+U can quantitatively reproduce the results from high level DFT calculations (PBE+U and HSE). The charge difference between $Fe_{Oct}^{3+}$ and $Fe_{Oct}^{2+}$ obtained by DFTB+U is around 0.65 e, which is larger than that obtained from PBE+U and HSE (around 0.35 e). As shown in Table 3, the atomic magnetic moment of $Fe^{2+}$ from DFTB+U agrees well with that by PBE+U and HSE. However, the magnetic moment of $Fe^{3+}$ is slightly overestimated especially compared with that from PBE+U calculations. This is consistent with the larger charge difference $\Delta Q$ obtained by DFTB+U. Note that for DFTB+U calculations, the total magnetic moments for the DBT and SCV surfaces have been fixed to 88 $\mu_B$ and 64 $\mu_B$, according to the values obtained with PBE+U method.[34]

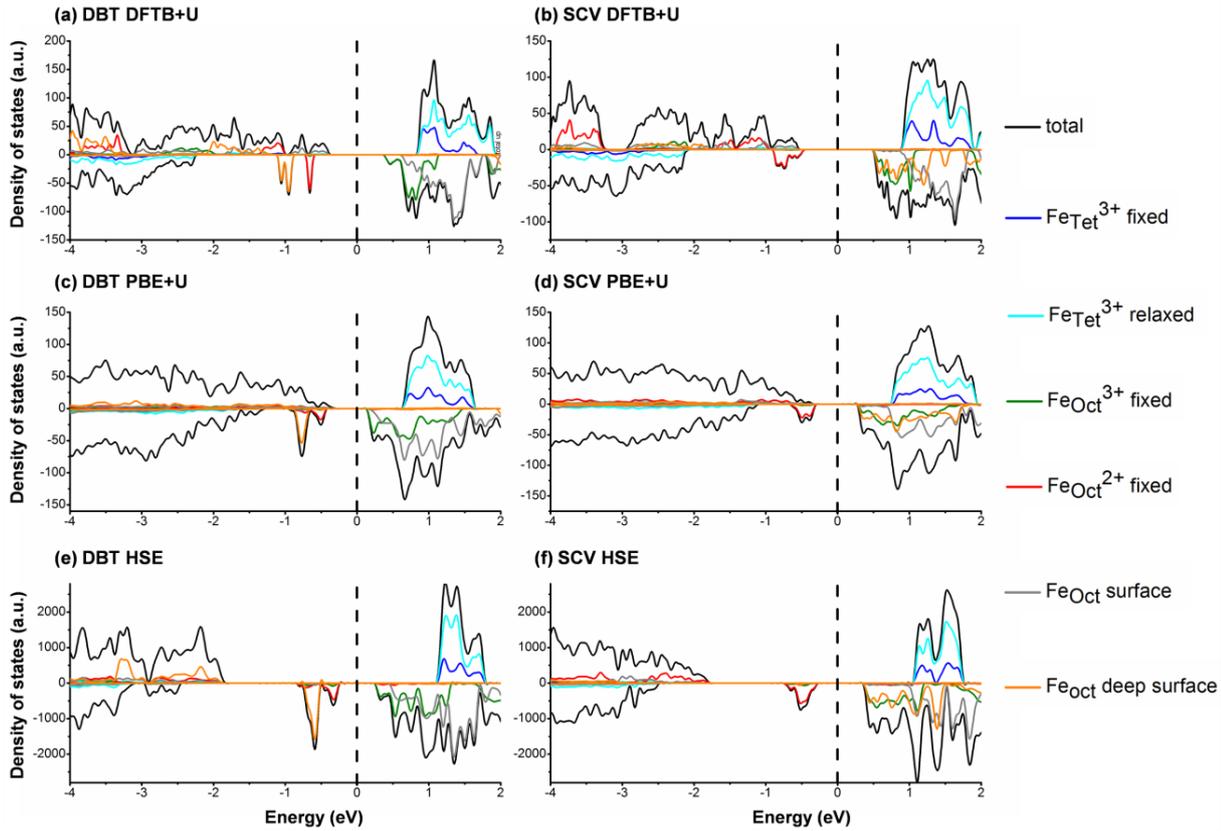

**Figure 4.** PDOS on the d states of different Fe ions in the SCV and DBT slab models. (a) and (b) are calculated using DFTB+U; (c) and (d) are calculated using PBE+U[34]; (e) and (f) are calculated using HSE[34]. The legend of colors is on the right. The black lines represent the total DOS. The blue and cyan lines represent PDOS on the $d$ states of $Fe_{Tet}^{3+}$ in the fixed layers and relaxed layers, respectively. The green and red lines represent PDOS on the $d$ states of $Fe_{Oct}^{3+}$ and $Fe_{Oct}^{2+}$ in the fixed layers. The gray lines represent PDOS on the $d$ states of $Fe_{Oct}$ in the surface layers (layer1 + layer3 + layer15 + layer17). The orange lines represent PDOS on the $d$ states of $Fe_{Oct}$ in the deep surface layers (layer5 + layer13). The Fermi level is scaled to zero as indicated by the dashed black lines.

The PDOS on the $d$ states of different Fe ions for both SCV and DBT surface models are calculated by DFTB+U and shown in Figure 4a, b. PDOS calculated by PBE+U and HSE from a previous work by some of us[34] are also plotted here (Figure 4c-f) for comparison. In general, the PDOS from DFTB+U agrees well with those from PBE+U and HSE. Both DBT and SCV surfaces are calculated to be semiconductors with a band gap between the $t_{2g}$ states from $Fe_{Oct}^{2+}$ and $Fe_{Oct}^{3+}$ in the inner bulk-like layers. The band gap values by DFTB+U are 0.89 eV and 0.85 eV for DBT and SCV surfaces, respectively, which are slightly larger than those by PBE+U (0.60 eV and 0.61 eV, respectively) and by HSE (0.56 eV and 0.69 eV, respectively). The characteristic peak in spin down states at about -0.5 eV observed both in experimental photoemission spectroscopy[69,70,71] and in previous computational studies[34] is also well reproduced. Considering the valence bands of the PDOS, DFTB+U is closer to PBE+U than to HSE. Based on the results just discussed above, we may conclude that DFTB+U can also rather satisfactorily describe not only the structural but also the electronic properties of magnetite surfaces.

## IV. CONCLUSION

In summary, in this work we have proposed a new DFTB parametrization of the Fe–O interactions based on previous Slater-Koster files, which enables it to describe rather accurately magnetite bulk and surface systems. The performance of DFTB+U with the proposed parameters is well assessed by investigating both magnetite bulk and (001) surface and through comparison with available experimental data and previous results by more sophisticated DFT methods (DFT+U and hybrid functional HSE).[34,44]

We observe that the lattice parameter of bulk magnetite can be calculated precisely by DFTB+U with an error of only 1.1% compared with the experimental value. The $Fe^{2+}/Fe^{3+}$ charge disproportionation in bulk magnetite cannot be reproduced due to the high symmetry. Nevertheless, the electronic properties of bulk magnetite obtained by DFTB+U agree well with those obtained by DFT+U and HSE with symmetry constrained wavefunction.

For the (001) surface, both the structural and electronic properties obtained by DFTB+U are in quantitative agreement with those obtained by DFT+U and HSE, including the charge disproportionation and charge distribution.

The excellent performance of DFTB+U on magnetite bulk and surfaces, in comparison to more sophisticated techniques, provides an efficient compromise between computational cost and accuracy for their simulation and, more interestingly, for the simulation of realistically large magnetite nanostructures. For example, with DFTB+U we could perform the full atomic relaxation of a magnetite nanocube containing about 1500 atoms with the edge length of 2.3 nm. Therefore, DFTB+U approach can handle magnetite nanoparticle of realistic size, whose simulation would be prohibitive with HSE and PBE+U methods. This work paves the way to innovative quantum mechanical studies on magnetite based nanomaterials, which attract increasing attention in medical applications.

**AUTHOR INFORMATION**

**Corresponding Author**

*E-mail: cristiana.divalentin@unimib.it

**ACKNOWLEDGMENT**

The authors are grateful to Lorenzo Ferraro for his technical help. The project has received funding from the European Research Council (ERC) under the European Union's HORIZON2020 research and innovation programme (ERC Grant Agreement No. 647020).